# Ranked Document Retrieval in (Almost) No Space

Nieves R. Brisaboa[1], Ana Cerdeira-Pena[1], Gonzalo Navarro[2] and Óscar Pedreira[1]

[1] Database Lab., Univ. of A Coruña, Spain.
{brisaboa,acerdeira,opedreira}@udc.es
[2] Dept. of Computer Science, Univ. of Chile. gnavarro@dcc.uchile.cl

**Abstract.** Ranked document retrieval is a fundamental task in search engines. Such queries are solved with inverted indexes that require additional 45%-80% of the compressed text space, and take tens to hundreds of microseconds per query. In this paper we show how ranked document retrieval queries can be solved within tens of milliseconds using essentially *no extra space* over an in-memory compressed representation of the document collection. More precisely, we enhance wavelet trees on bytecodes (WTBCs), a data structure that rearranges the bytes of the compressed collection, so that they support ranked conjunctive and disjunctive queries, using just 6%–18% of the compressed text space.

## 1 Introduction

Ranked document retrieval on a large collection of natural-language text documents is the key task of a search engine. Given a query composed of a set of terms, the engine returns a list of the documents most relevant to the query. Typical relevance measures include *tf-idf* and Okapi BM25.

Efficient ranked document retrieval relies on the use of inverted indexes [1,2,3], which maintain for each term of the vocabulary a posting list with the identifiers of the documents in which that term occurs. Given a query, the system computes the union (*bag-of-words* query) or intersection (*weighted conjunctive* query) of the posting lists of the terms composing the query, keeping only the documents with highest relevance with respect to the query.

The inverted index does not support by itself all the operations needed in a complete search engine. For example, a search engine not only returns the ranked list of documents, but it usually shows a snippet for each result, or even offers a cached version of the document. This requires storing the text of the documents in addition to the index. Compressing the text data and the inverted index is useful not only to save space. On disk-based systems, it reduces the amount of I/O needed to answer queries. A recent trend (e.g., [3,4,5,6]) is to maintain all the data in main memory, of a single machine or a cluster (in the case of large search engines). In this case, compression helps keeping larger collections in main memory, or using fewer computers and less energy in the case of clusters.

The texts of the documents are usually stored in a compressed form that allows fast decompression of random portions of the text. Such compressors achieve 25%–30% of the size of the original text. Inverted indexes are also compressed, and amount to an additional 15%–20% of the size of the original text, or 45%–80% of the size of the compressed text [1,2,7,8]. Typical query times of in-memory systems are in the orders of tens to hundreds of microseconds.

A recent alternative to storing the text plus the inverted index is the *Wavelet Tree on Bytecodes (WTBC)* [9,10]. This data structure reorganizes the bytes of the codewords output by any word-based byte-oriented semistatic statistical compressor. Within the space of the compressed text (i.e., around 30%–34% of the text size) the WTBC not only can extract arbitrary text snippets or documents, but it also solves *full-text* queries, that is, it finds the exact positions where a word or phrase occurs in the collection. Full-text queries are usually solved with a *positional* inverted index, which stores exact word positions, yet this is outperformed by the WTBC when little space over the compressed text is available. The representation was later extended to *document retrieval* queries, that is, listing all the distinct documents where a query appears [11]. However, *ranked document retrieval* queries, which find only a few most relevant documents for the query and are arguably the most important ones from the point of view of the end-user, have not been addressed under this scheme.

In this paper we close this gap. We show how WTBCs can be extended to efficiently support ranked document retrieval queries as well. As a result, all the main IR queries can be carried out on top of a data structure that requires just 6%-18% on top of the compressed text space (2.0%–5.5% of the original text space). The times of the WTBC to solve ranked document retrieval queries are in the order of milliseconds, which is significantly higher than inverted index times. However, those times are still reasonable in many scenarios and the solution offers important space advantages compared to the 45%–80% of extra space posed by inverted indexes.

For example, consider the case of a fully-functional search engine. A classical approach could compress the text to 25% of its original size (using word-based bit-Huffman to give direct access to the text), plus 15%-20% for a document retrieval index. If precise word occurrences are to be spotted in order to display a snippet around them, then a positional inverted index is necessary, which requires 25% further space. All this functionality is offered by the enhanced WTBC, which requires around 32%–39.5% of the original text space (30%–34% for a byte-oriented compressor plus 2.0%–5.5% of redundancy). Dropping from 65%–70% to 32%–39.5%, that is, halving the space, may be key to avoid using secondary storage, to using fewer machines, or even to achieve a feasible solution when the memory is limited (as in mobile devices).

## 2  WTBC: Wavelet Trees on Bytecodes

The *Wavelet Tree on Bytecodes* (WTBC) [9,10] is a method for representing natural language texts in a compressed form, so that random access to any portion of the text, and search for the occurrences of any term, are efficiently supported. The WTBC is built on a text compressed using any word-based byte-oriented semistatic statistical compressor, by rearranging the codewords into a wavelet-tree-like [12] structure.

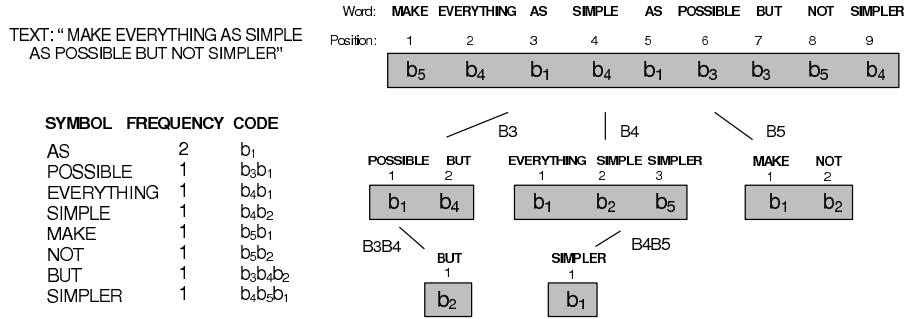

**Fig. 1.** Example of a WTBC.

### 2.1 Document Compression

Classical statistical compression methods model the text as a sequence of characters and assign shorter codewords to more frequent characters, where each codeword is a sequence of bits. Huffman coding [13] is the best-known example, although many more have been proposed.

On natural language texts, it has been shown that using words, not characters, as the source symbols, significantly improves compression ratios. On the other hand, decompression time can be boosted by using byte-oriented encoders, where codes are sequences of bytes instead of bits. This yields compression ratios around 30%–34%. See any recent work for references, e.g., [14].

Dense codes (or $(s,c)$-DC) [14] are particularly convenient for our work. The vocabulary words are sorted by decreasing frequency and encoded as a sequence of bytes. The codewords are formed by zero or more *continuers* (byte values in the range $[s..s+c)$) terminated by one *stopper* (a byte value in the range $[0..s)$), where $s+c = 256$. Then the $s$ most frequent words are assigned 1-byte codewords, the next $sc$ are assigned two-byte codewords, the next $sc^2$ are assigned three-byte codewords, and so on. The pair $(s,c)$ is chosen so as to optimize compression.

### 2.2 Reorganizing Compressed Text

The basic idea in the WTBC is to rearrange the text by placing the different bytes of each codeword in different nodes of a tree. The root of the tree is an array containing the first byte of the codeword of each word in the text, in the same order they appear in the original text. The second byte of each codeword (having more than one byte) is placed on the second level of the tree. The root has one child per continuer value $B_b$. Each node $B_b$ in the second level contains the second bytes of the codewords of the words whose first byte is $b$, again following the same order of the source text. And so on for the next levels.

Figure 1 shows an example of a WTBC built for the text 'MAKE EVERYTHING AS SIMPLE AS POSSIBLE BUT NOT SIMPLER'. In the example, bytes $b_1$ and $b_2$ are stoppers, and $b_3$ to $b_5$ are continuers (not all the combinations are used).

The main operations in a WTBC are *decoding* the word at a given position of the text, *locating* the occurrences of a word, and *counting* the number of occurrences of a word. These algorithms are based on the use of *rank* and *select* operations over the bytemap of each node to determine the path to follow on the tree. Given a bytemap $B = \{b_1, \ldots, b_n\}$:

- $rank_b(B, i)$ = number of occurrences of byte $b$ in $B$ up to position $i$.
- $select_b(B, i)$ = position of the $i^{th}$ occurrence of the byte $b$ in bytemap $B$.

Partial counters are maintained for each bytemap in order to efficiently compute *rank* and *select*. These solve the queries within a few microseconds while posing just a 3% space overhead over the original text size [10].

In order to *decode* the word at a given position in the text we go down in the tree by using *rank* operations at each step. In the example of Figure 1, to decode the word at position 9, we start by reading $root[9] = b_4$. Since $b_4$ is a continuer, we move to the next level of the tree, to node $B4$. This node holds the second byte of all the codewords starting with $b_4$, following the order of the text. Thus, the next byte of the word we are decoding will be at position $rank_{b_4}(root, 9) = 3$ of node $B4$, that is $B4[3] = b_5$. Again, $b_5$ is a continuer, and $rank_{b_5}(B4, 3) = 1$ tells us that the third byte of that word will be in the node $B4B5$ at position 1. Finally, we obtain $B4B5[1] = b_1$, which is a stopper. As a result, we have the codeword $b_4 b_5 b_1$, which corresponds to the word 'SIMPLER'.

For *locating* the occurrences of a word we traverse the tree from a leaf to the root. For example, assume we want to find the first occurrence of the word 'BUT'. In Figure 1, its codeword is $b_3 b_4 b_2$, so the search starts at node $B3B4$, where we locate the first occurrence of $b_2$ by computing $select_{b_2}(B3B4, 1) = 1$. Hence the first position at node $B3B4$ corresponds to the first occurrence of 'BUT'. Now we find the position of the first $b_4$ in node $B3$, which is $select_{b_4}(B3, 1) = 2$. Thus our codeword is the second one starting by $b_3$ in the root node. Finally, we find the position of the second $b_3$ in the root of the tree, $select_{b_3}(root, 2) = 7$. Thus the first occurrence of 'BUT' is at the $7^{th}$ position of the text.

In order to *count* the number of occurrences of a word, we find the node in the WTBC where the last byte of its codeword appears. In that node we can efficiently count the number of occurrences of the word by counting how many times the last byte of the codeword appears using *rank* on the bytemap. We can also *count* the occurrences within a range of the text, by tracking down the range endpoints toward the leaf of the word, and computing the difference of *rank* values on the mapped range.

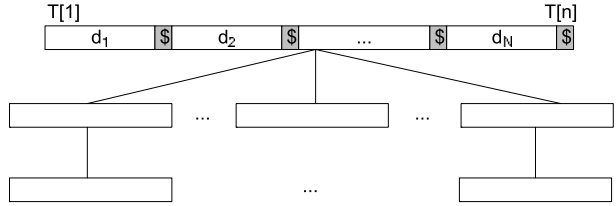

**Fig. 2.** WTBC over a collection of concatenated documents.

## 3 Efficient Ranked Document Retrieval

In this section we present our proposal for solving ranked document retrieval queries using the WTBC over $(s,c)$-DC. We concatenate all documents of the corpus in a single text string. We assume that each document ends with a special symbol '$\$$', which then becomes a document separator (just as in [11]). Then, the string is compressed with $(s,c)$-DC and a WTBC is built on the result of the compression (see Figure 2). For efficiency reasons, we reserve the first codeword of the $(s,c)$-DC encoding scheme for the '$\$$' symbol, so the document separator can be easily found in the root of the tree, since its codeword has only one byte.

We consider top-$k$ *conjunctive* (AND) and *bag-of-words* (OR) queries, in which only the $k$ most relevant documents for the query must be returned. We present two different alternatives. A first one, using no extra space over the WTBC, computes the union or the intersection prioritized by relevance, so the documents are found in decreasing relevance order and we stop as soon as we find $k$ of them. A second one, using small additional bitmaps, first computes the union or the intersection and then collects the $k$ most relevant documents.

In this paper we use the *tf-idf* relevance measure. For a query word $w$ and a document $d$, this is $tf_{w,d} \cdot idf_w$, where $tf_{w,d}$ is the number of occurrences of $w$ in $d$, and $idf_w = \log N/df_w$, where $N$ is the number of documents in the collection and $df_w$ is the number of documents where $w$ appears. Values $df_w$ are stored, one per word, in our index within insignificant extra space, as the vocabulary size becomes irrelevant as the collection grows [15]. For a query of several words, the *tf-idf* relevance of a document is the sum of that of the query words.

### 3.1 Solution with No Extra Space (WTBC-DR)

The procedure uses a priority queue storing *segments*, that is, concatenations of consecutive documents. The priority will be given by the *tf-idf* value of the concatenations (seen as a single document). We start by inserting in the queue the segment that corresponds to the concatenation of all the documents, with its associated priority obtained by computing its *tf-idf* value. A segment is represented by the corresponding endpoints in the root bytemap $T[1,n]$ of the WTBC. Since the *idf* of each word is precomputed, to compute *tf-idf* relevance value we only need to calculate the *tf* of each word in the segment, that is, we *count* its number of occurrences in the segment, as explained at the end of Section 2.2.

**Algorithm 1:** ranked bag-of-words retrieval with WTBC-DR

**Input**: $wt$ (WTBC), $q$ (query), $k$ (top-$k$ limit)
**Output**: list of top-$k$ ranked documents
$s.start\_pos \leftarrow 1$; $s.end\_pos \leftarrow n$; $s.score \leftarrow tfidf(s, q)$; $s.ndocs \leftarrow N$;
$pq \leftarrow \langle\rangle$; $insert(pq, s)$ // $s.score$ is the priority for queue $pq$ ;
**while** *less than $k$ documents output* **and** $\neg empty(pq)$ **do**
    $s \leftarrow pull(pq)$;
    **if** $s.ndocs = 1$ **then**
        output $s$
    **else**
        $\langle s_1, s_2 \rangle \leftarrow split(s)$ // computes $s_i.start\_pos$, $s_i.end\_pos$ and $s_i.ndocs$;
        $s_1.score \leftarrow tfidf(s_1, q)$; $s_2.score \leftarrow s.score - s_1.score$;
        $insert(pq, s_1)$; $insert(pq, s_2)$;
    **end**
**end**

The procedure repeatedly extracts the head of the queue, that is, the segment with the highest priority (the first time, we extract the segment $T[1, n]$). After extracting a segment, the procedure checks if it corresponds to just a single document, or to more than one.

If the extracted segment has more than one document, the procedure splits it into two subsegments, by using the '$' symbol closest to the middle of the segment, as the point to divide it. This '$' is easily found using $rank$ and $select$ on $T$ (i.e., for a segment $T[a, b]$, we use, roughly, $select_\$(T, rank_\$(T, (a + b)/2)))$, which also tell us the number of documents in each subsegment. After the division, the relevance of each of the two subsegments is computed, and they are inserted in the queue using their relevance as priority. Note that it is sufficient to compute the *tf-idf* of one subsegment (using *count* of each query word in the subsegment to compute the *tf* values), and then the *tf-idf* value of the other subsegment is found by subtraction from the main segment's *tf-idf*.

If, instead, the extracted segment contains only one document, it is directly output (with its *tf-idf* relevance value), as the next most relevant document. This is correct because *tf-idf* is monotonic over the concatenation: the *tf-idf* of the concatenation of two documents is not smaller than the *tf-idf* of any of them. Thus the relevance of the individual document extracted is not lower than that of any other remaining in the priority queue.

The iterative process continues until we have output $k$ documents. In this way, it is not necessary to process all the documents in the collection, but rather the search is guided towards the areas that contain the most promising documents for the query until it finds the top-$k$ answers. Note that the procedure does not need to know $k$ beforehand; it can be stopped at any time.

The pseudocode for bag-of-words queries is given in Algorithm 1. For weighted conjunctive queries we add an additional check during the procedure: if a segment does not contain some of the words in the query (i.e., some *tf* is zero), the segment is discarded without further processing.

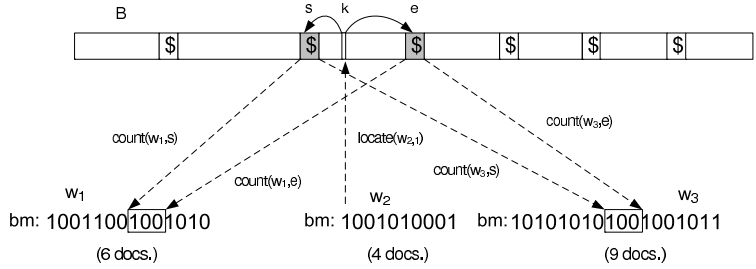

**Fig. 3.** Search step for conjunctive queries using additional bitmaps.

### 3.2 With Additional Bitmaps (WTBC-DRB)

This solution uses an additional bitmap for each word in the vocabulary whose *idf* exceeds a fixed threshold $\epsilon$.[1] The bitmap of each word encodes the number of documents where it appears and its term frequency in each of those documents, as follows: a 1 in the bitmap marks a new document where the word appears, and it is followed by as many 0s as the number of occurrences of the word in that document, minus one. For example, if the bitmap of a word is 10000100100000, it expresses that the word appears in three documents, with five occurrences in the first one, three in the second, and six in the third.

**Conjunctive Queries** This time we will obtain all the documents in the intersection, and choose the top-$k$ from those. Figure 3 shows an example of the search using bitmaps. Let us consider this example to explain how the algorithm for conjunctive queries works. It assumes a query composed of three words $w_1$, $w_2$, and $w_3$, that appear in 6, 4, and 9 different documents, respectively.

The algorithm will insert the documents in the intersection in a *result* priority queue sorted by *tf-idf* relevance and with size limited to $k$. At the end, this queue will contain the answers. It will also compute triplets $(wID, nDocs, i)$ for each query word: $wID$ is the identifier of the word, $nDocs$ is the number of documents that have not been yet processed for that word, and $i$ is the position in the bitmap of the 1 marking the first document not yet processed for that word. At each step, it will process the triple with lowest $nDocs$ value, that is, the query word with least documents to be processed.

In the example, the initial triplets are $\langle (w_2, 4, 1), (w_1, 6, 1), (w_3, 9, 1) \rangle$. Word $w_2$ will be processed first since it has the lowest number of documents to be processed, 4. In each step, if we chose triplet $(wID, nDocs, i)$, we locate the $i^{th}$ occurrence of $wID$ (note that the bits in the bitmap of a word correspond to its occurrences in the text). Then, we identify in which document it is contained and also the document limits. In the example, we choose $(w_2, 4, 1)$. Since $i = 1$, we locate the first occurrence of $w_2$ in the root $T$ by traversing the tree from the leaf of *wID* (see Section 2). Let $p$ be the position found in $T$. Then the identifier of the document containing that occurrence of $w_2$ is $d = 1 + rank_\$(T, p)$. To find the beginning and the end ($s$ and $e$, in Figure 3) of document $d$, we can use $s = select_\$(T, d-1)$ and $e = select_\$(T, d)$.[2]

---

[1] If the *idf* of a word is lower than this threshold, we assume it is not relevant for the global document scores. The idea is to filter out stopwords.

Once the document limits have been obtained, we *count* the occurrences of the other query words restricted to the limits of this document (recall the end of Section 3.1). If some of the words does not appear in the document, it is discarded. If all the query words appear in the document, we compute its *tf-idf* using the term frequencies just computed (i.e., the counts). The *tf* value for word *wID* can be computed more efficiently without traversing the wavelet tree, but using the corresponding bitmap ($bm_{w_2}$, in the example). We only have to search for the position of the next 1 in its bitmap (which can be done in constant time [16]), and compute the *tf* as the difference between that position and $i$. In the example, where $i = 1$, as mentioned before, the next 1 of $bm_{w_2}$ is at position 4, thus the *tf* value for $w_2$ is $4 - 1 + 1 = 4$. Finally, the computed *tf-idf* value is used to insert the document in the results queue.

Be the document inserted or discarded, the procedure continues by considering the documents to the right of the current one, that is, by regarding the documents of each word's bitmap to the right of the bit corresponding to the document just processed. This is done by recomputing the triplets. For each word of the query, the information provided by the *count* operations performed over the WTBC to obtain its *tf* value in the document (or on the bitmap, in the case of the word *wID*) is mapped to its corresponding bitmap to recompute its triplet. For example, consider $w_1$. Its number of occurrences before $s$ is $count(w_1, s) = 7$, and before $e$, $count(w_1, e) = 10$. Then, the $i$ value in the new triplet for word $w_1$ is $count(w_1, e) + 1 = 11$. The new *nDocs* value can also be easily computed with a constant-time *rank* operation [16] on its bitmap $bm_{w_1}$: $df_{w_1} - rank_1(bm_{w_1}, count(w_1, e)) = 6 - 4 = 2$. This means that $w_1$ still appears in $nDocs = 2$ documents after the document we have just processed. By proceeding in the same way with the rest of the words, we obtain the three new triplets: $\langle (w_1, 2, 11), (w_2, 3, 4), (w_3, 3, 12) \rangle$.

The search proceeds iteratively until there are no documents to be processed for some of the words, that is, until we reach a triplet where $nDocs = 0$.

**Bag-of-words Queries** These queries do not benefit from the use of bitmaps as much as conjunctive queries, since every document where any of the query terms appears must be considered. In that case, we proceed as follows for each word, $w_j$, of the query. We sequentially locate in the WTBC the first occurrence of $w_j$ in each new document where it appears, by traversing the 1s of its bitmap (jumping to the next 1 in constant time [16]). For each new 1, say found at bitmap position $p$, we locate the $p$-th occurrence of $w_j$ in the WTBC. Then the identifier of that document is found using $rank_\$$ on $T$, and the term frequency of $w_j$ in such document is simply the distance to the next 1 in the bitmap. All the documents where $w_j$ appears, with its frequencies, are stored in a set. Once we have aggregated all the documents where all the query words appear, we sort it by document identifier, add up the contributions to *tf-idf*, and choose the top-$k$.

---
[2] In practice we use faster structures for those particular cases of *select*, where one wants the preceding and following occurrences of '$'.

**Table 1.** Description of the corpus used and compression properties

| Corpus | size (MB) | #docs | #words | voc. size |
|---|---|---|---|---|
| ALL | 987.43 | 345,778 | 219,255,137 | 718,691 |

| Technique | CR | CT | DT |
|---|---|---|---|
| WTBC-DR | 35.0 | 40.1 | 8.6 |
| WTBC-DRB | 38.0 | 65.6 | 8.5 |

## 4 Experimental Evaluation

We evaluated the performance of the proposed algorithms over a data set created by aggregating text collections from TREC-2: AP Newswire 1988, and Ziff Data 1989-1990, as well as TREC-4, namely Congressional Record 1993, and Financial Times 1991 to 1994. All of them form a document corpus of approximately 1GB (ALL). Table 1 (left) gives the main statistics of the collection used: size in MBytes, number of documents, total number of words (using the spaceless word model [17]), and number of different words (vocabulary size).

We ran the experiments in a system with an AMD Phenom II X4 955 Processor (3.2 GHz) and 8GB RAM. It ran Ubuntu 9.10 GNU/Linux (kernel version 2.6.31.23). The compiler used was gcc version 4.4.1 and `-O9` compiler optimizations were applied. Time results measure CPU user time in seconds.

### 4.1 Document Representation

Table 1 (right) shows the results obtained when the collection is represented with WTBC over $(s,c)$-DC, and give the compression ratio (CR) (in % of the size of the original text collection), as well as the time to create the structures (CT) and to recover the whole text back from them (DT), in seconds, for the two variants we have proposed. Notice that the difference between WTBC-DR and WTBC-DRB is given by the storage of the words' bitmaps in the second one. The raw compressed data uses around 32.5% of the space used by the plain text, and the WTBC requires an additional waste of 2.5% of extra space for the bytemap *rank* and *select* operations, for a total of 35%. Structure WTBC-DRB adds an additional 3% of space (we have used $\epsilon = 10^{-6}$ for the bitmaps, which leaves out just 65 words, mostly stopwords), for a total of 38%.

### 4.2 Ranked Document Retrieval

Tables 2 and 3 show the average times (in milliseconds) to find the top-$k$ (using $k = 10$ and $k = 20$) ranked documents for a set of queries, using WTBC-DR and WTBC-DRB techniques. We considered different sets of queries. First, we generated synthetic sets of queries, depending on the document frequency of the words ($f_{doc}$): *i)* $10 \leq f_{doc} \leq 100$, *ii)* $101 \leq f_{doc} \leq 1,000$, *iii)* $1,001 \leq f_{doc} \leq$ 10,000 , and $iv)$ 10,001 $\leq f_{doc} \leq$ 100,000, and also on the number of words that compose a query, namely, 1, 2, 3, 4 and 6. Each set is composed of 200 queries of words randomly chosen from the vocabulary of the corpus, among those belonging to a specific range of document frequency. Second, we also used queries from a real query log[3] (*real*), and created 5 sets of 200 queries randomly chosen composed of 1, 2, 3, 4, and 6 words, respectively. The same sets of queries were used for dealing with both conjunctive (Table 2) and bag-of-words scenarios (Table 3).

**Table 2.** Results for top-10 and top-20 1-word queries and conjunctive queries

| $f_{doc}$ | Technique | #words per query | | | | | | | | | |
|---|---|---|---|---|---|---|---|---|---|---|---|
| | | 1 | | 2 | | 3 | | 4 | | 6 | |
| | | top-10 | top-20 | top-10 | top-20 | top-10 | top-20 | top-10 | top-20 | top-10 | top-20 |
| $i)$ | WTBC-DRB | 0.38 | 0.37 | 0.35 | 0.35 | 0.27 | 0.27 | 0.25 | 0.25 | 0.23 | 0.22 |
| | WTBC-DR | 2.27 | 3.45 | 0.86 | 0.87 | 0.55 | 0.55 | 0.43 | 0.42 | 0.28 | 0.28 |
| $ii)$ | WTBC-DRB | 4.20 | 4.20 | 5.13 | 5.14 | 4.45 | 4.45 | 4.09 | 4.09 | 3.61 | 3.60 |
| | WTBC-DR | 6.18 | 7.80 | 9.57 | 9.61 | 6.54 | 6.55 | 4.70 | 4.71 | 3.44 | 3.44 |
| $iii)$ | WTBC-DRB | 23.34 | 23.38 | 33.70 | 33.70 | 27.87 | 27.91 | 24.09 | 23.95 | 21.43 | 21.44 |
| | WTBC-DR | 15.06 | 18.62 | 63.05 | 72.37 | 64.06 | 66.67 | 53.51 | 53.63 | 44.19 | 46.31 |
| $iv)$ | WTBC-DRB | 191.34 | 191.78 | 279.66 | 279.65 | 263.79 | 263.64 | 240.16 | 240.49 | 207.65 | 207.91 |
| | WTBC-DR | 53.40 | 66.15 | 151.16 | 185.76 | 284.92 | 341.42 | 382.92 | 415.76 | 404.26 | 410.35 |
| real | WTBC-DRB | 19.14 | 19.17 | 29.67 | 29.66 | 39.95 | 39.94 | 34.38 | 34.36 | 33.60 | 33.60 |
| | WTBC-DR | 6.68 | 9.08 | 34.92 | 41.55 | 67.36 | 77.52 | 78.34 | 87.11 | 101.22 | 108.95 |

**Conjunctive Queries** WTBC-DRB is, in general, faster than WTBC-DR. This shows that first computing the intersection and then ranking it is a good strategy compared to trying to prioritize the intersection by the relevance of segments, especially if the former strategy can be sped up with bitmaps. The situation is reversed when the queries have low selectivity (see the values of 1-word queries, for $iii)$, $iv)$, and *real*; and the results of 2-word queries, for case $iv)$). In those situations, the amount of documents in the intersection is presumably quite large (in the case of 1-word queries, this value is precisely given by $f_{doc}$), thus WTBC-DRB must process each of them, whereas WTBC-DR can still benefit from processing first the most promising documents, and stopping when the first $k$ are retrieved.

For both techniques, the processing times decrease as the number of words in the query increases, within a given $f_{doc}$ band. This is expected when the words are chosen independently at random, since more words give more pruning opportunities for both algorithms. However, in the scenario $iv)$, where words appear in too many documents, and in *real*, where the query words are not independent, the WTBC-DR pruning is not efficient enough and its times grow with the number of words. For *real* queries, WTBC-DRB is non-monotonic, worsening up to 3 words and then dropping slowly.

---

[3] Obtained from TREC (http://trec.nist.gov/data/million.query.html)

Table 3. Results for top-10 and top-20 bag-of-words queries

| $f_{doc}$ | Technique | #words per query | | | | | | | |
|---|---|---|---|---|---|---|---|---|---|
| | | 2 | | 3 | | 4 | | 6 | |
| | | top-10 | top-20 | top-10 | top-20 | top-10 | top-20 | top-10 | top-20 |
| $i)$ | WTBC-DRB | 23.94 | 23.85 | 24.40 | 24.25 | 24.72 | 25.03 | 25.73 | 25.70 |
| | WTBC-DR | 3.86 | 4.91 | 4.28 | 5.91 | 5.17 | 7.35 | 6.86 | 9.53 |
| $ii)$ | WTBC-DRB | 31.84 | 31.93 | 36.43 | 36.25 | 41.13 | 41.20 | 49.52 | 49.56 |
| | WTBC-DR | 10.12 | 13.74 | 14.22 | 19.74 | 18.56 | 24.53 | 27.78 | 35.91 |
| $iii)$ | WTBC-DRB | 72.06 | 72.15 | 97.81 | 97.92 | 120.67 | 121.04 | 167.55 | 167.42 |
| | WTBC-DR | 29.63 | 38.54 | 43.20 | 57.81 | 61.65 | 78.06 | 94.96 | 118.71 |
| $iv)$ | WTBC-DRB | 384.11 | 384.06 | 585.99 | 585.81 | 770.20 | 770.95 | 1,142.62 | 1,143.97 |
| | WTBC-DR | 98.84 | 125.52 | 156.79 | 202.71 | 223.31 | 281.07 | 359.84 | 462.16 |
| $real$ | WTBC-DRB | 129.46 | 129.43 | 263.85 | 263.82 | 372.54 | 372.20 | 686.76 | 687.04 |
| | WTBC-DR | 27.96 | 36.17 | 58.91 | 75.74 | 82.80 | 106.42 | 150.94 | 192.02 |

**Bag-of-words Queries** In this scenario, the more query words, the higher is the average processing time in both alternatives, since each word increases the number of valid documents. Unlike the previous scenario, WTBC-DR beats WTBC-DRB, as in this context one can hardly profit from the bitmaps. In WTBC-DRB, a sequential processing of the documents where each word appears is necessary, whereas WTBC-DR processes only the most promising segments.

In the case of real queries, the processing time also increases in terms of the number of words for both techniques, independently of the document frequency of the words composing the query. The results show that WTBC-DR performs better than WTBC-DRB, just as on synthetic queries.

## 5 Conclusions

We have shown how the Wavelet Tree on Bytecodes (WTBC), a compressed data structure that supports full-text searching and document retrieval within essentially the space of the compressed text, can be enhanced to support also ranked document retrieval, which is by far the most important operation in IR systems, within tens of milliseconds. The enhanced WTBC becomes a very appealing solution in scenarios where minimizing the use of main memory is of interest, as it supports all the typical repertoire of IR operations at basically no storage cost. In particular it may help maintaining an indexed collection entirely in RAM when a classical solution would have to resort to disk, or reduce the number of computers needed for a cluster that implements a large in-memory distributed index, or allow implementing an indexed collection on a mobile device with limited memory, just to mention some obvious application scenarios.

We have presented two proposals. A first one does not use any extra space on top of the WTBC, and solves bag-of-words (disjunctive) queries and weighted conjunctive queries within milliseconds. The second uses a very small amount of extra space, and noticeably speeds up selective conjunctive queries. Another advantage of our second strategy is that it easily generalizes to other weighting schemes beyond *tf-idf*, for example Okapi BM25, because it simply computes the relevance of all the candidates and then chooses the best ones. The first scheme applies a more sophisticated prioritized traversal, which is not so easy to adapt to the BM25 weighting formula, as it includes factors like the document size.

We plan to carry out experiments on larger collections for the full paper. It is quite clear, however, that even if the times grew linearly with the collection size (which is expected on the second technique but is probably pessimistic on the first), we would still have decent response times on 100GB collections, which are on the limit of what can be handled in main memory on today's machines. Another aspect worth of further study is the possibility of reducing the search times with inexact top-$k$ document retrieval. While heuristics that may miss some relevant answers are widely used in inverted-index-based algorithms to improve performance, we have not allowed ourselves to profit from those.